\documentclass[aps,pre,preprint,superscriptaddress]{revtex4-1}

\usepackage{amsmath,amssymb,graphicx}
\usepackage{enumitem}
\usepackage{hyperref}
\hypersetup{colorlinks=true, linkcolor=red, citecolor=blue, urlcolor=blue}
\usepackage{amsbsy}
\usepackage{latexsym}
\usepackage{color}
\usepackage{graphicx}
\usepackage{psfrag}
\usepackage[normalem]{ulem}
\usepackage{bm}
\usepackage{lipsum}
\usepackage{color}
\usepackage{tikz}
\usepackage{float}
\usepackage{multirow}

\newcommand{\be}{\begin{equation}}
\newcommand{\ee}{\end{equation}}
\newcommand{\bea}{\begin{eqnarray}}
\newcommand{\eea}{\end{eqnarray}}

\newcommand{\comment}[1]{}

\begin{document}

\title{Self-assembly of Dipolar Crystals from Magnetic Colloids}

\author{Anuj Kumar Singh}
\email{anuj0630@gmail.com}
\affiliation{Department of Physics, Indian Institute of Technology Delhi, New Delhi, 110016, India}

\author{Sanjay Puri}
\email{purijnu@gmail.com}
\affiliation{School of Physical Sciences, Jawaharlal Nehru University, New Delhi 110067, India }

\author{Varsha Banerjee}
\email{varsha@physics.iitd.ac.in}
\affiliation{Department of Physics, Indian Institute of Technology Delhi, New Delhi, 110016, India}

\begin{abstract}
We study the self-assembly of magnetic colloids using the Stockmayer (SM) model characterized by short-range Lennard-Jones interactions and long-range dipole-dipole interactions. Using molecular dynamics simulations, we design cooling protocols that yield perfectly assembled single-domain magnetic crystals. We identify cooling rates at which the system transforms from an amorphous glass to a crystal, where magnetic ordering promotes crystalline order. Remarkably, we observe that the latter develops via a spontaneous transition rather than through the traditional nucleation and growth mechanism. For a weakly dipolar fluid ($\mu=1$), this self-assembly results in a face-centered cubic (FCC) colloidal crystal with dipole moments chained along the (111) direction. For fluids with higher dipole moment ($\mu = 2.5$), the crystal structure shifts towards a body-centered orthorhombic (BCO) arrangement due to the compression of chains from strong dipolar attractions. These results provide valuable insights into the mechanisms driving crystallization in magnetic fluids, opening new avenues for understanding the formation of magnetically responsive colloidal magnetic crystals with promising applications.
\end{abstract}

\maketitle

\section{Introduction}

Self-assembly is the auto-organization of the constituent particles in a colloidal system from a disordered state to an ordered state without any external intervention \cite{Whitesides2002, Snezhko2011, Abelmann2020, Sacanna2013, Messina2014}. The driving force is usually thermodynamic, and the self-assembled structures have a lower free energy. However, this does not guarantee that such a structure is in equilibrium. Many times, it can be far from equilibrium with no trace of the global minimum. Some systems, when rapidly cooled to very low temperatures, can form amorphous glass. In this case, the motion of the constituents becomes so slow that the structure of the fluid {\it vitrifies} and arrests crystal formation \cite{Sanz2011, Tanaka2022, Koperwas2016}. These kinetically trapped structures occur when the component interactions are strong and the defects cannot be erased on the timescale of observation. Although customarily regarded as undesirable, many kinetically trapped structures have interesting and useful properties, examples being gels, diffusion-limited aggregates and structural glasses, to name a few \cite{Swan2012, Royall2021, Licinio2001, Krotzky2016}. Glasses need not be long-lived, as even those that appear permanent can unexpectedly crystallize in a process known as {\it devitrification} \cite{Ingebrigtsen2019, Ganapathi2021}. Theoretical guidance to a better mechanistic understanding of how glassiness can be preserved or nudged toward crystallization despite structural arrest can prove to be of significant value in material synthesis.

In the past, there have been many studies on self-assembly in nonmagnetic colloids \cite{Paci2005, Shneidman1998, Mausbach2020}. These have been widely modeled by the Lennard-Jones (LJ) potential, which includes repulsive and attractive terms. More recently, there has been great interest in magnetic colloids or {\it ferrofluids} that comprise magnetic dipoles dispersed in a carrier liquid \cite{Van1993, Vega2019, Gao2000, Groh1996}. This fascinating system exhibits a gas-liquid (GL) coexistence phase diagram that can be controlled by the strength of the magnetic moments of the constituent particles. The dual characteristics of fluidity and magnetism introduce a plethora of novel structures, both in the liquid state and in the solid state, which help to understand diverse complex phenomena. For example, in dilute fluids, the dipoles align to form long chains, which then become rings as the temperature is lowered \cite{Ivanov2015, Ivanov2022}. The ensuing entangled magnetic strings are a playground for understanding loopless branched structures, gels, polymerization, and the phase behavior of network fluids.  High-density magnetic fluids, on the other hand, offer a platform to study the self-assembly of magnetic crystals and arrested glassy states \cite{Zhang2020, Van2018}. They are prototypes for studying self-assembly pathways, temperature-induced structural transitions, heterogeneities in glass formers, etc. However, it is experimentally challenging to study the particle-scale dynamics in atomic or molecular systems because of the small size of their constituents. So colloidal systems, whose particles are visible by microscopy, have been widely used to explore the physics of glasses and crystallization \cite{Pham2002, Van1993change, Schope2006, Gasser2001}. 

Magnetic colloids are well described by the Stockmayer (SM) model which incorporates long-ranged anisotropic dipole-dipole interactions along with the usual short-ranged isotropic LJ interaction \cite{Van1993, Smit1989, Leeuwen1993, Bartke2007, Kalyuzhnyi2007, Dudowicz2004, Leeuwen1994, Hentschke2007}. It exhibits the GL coexistence observed in experiments and is a playground to comprehend the influence of the dipolar interactions during the formation of the magnetic condensates. In recent work, we studied the self-assembly of SM particles after performing quenches from the high-temperature homogeneous phase to lower temperatures in the coexistence curve \cite{Singh2023accelerated, Singh2023phase}. The condensates (droplets) exhibited density-dependent shapes and were composed of closed-packed chains of dipoles that imparted characteristic magnetic properties. For quenches much below the freezing temperature $T_f$, the aggregates were amorphous and glassy with no magnetic or long-range positional order. An important line of investigation then is to look for protocols that yield perfectly assembled magnetic crystals or preserve glassiness as the need may be. We undertake these tasks by performing extensive molecular dynamics (MD) simulations of the SM fluid and address the following questions: What are the different transitions an SM fluid undergoes during cooling? What are the cooling rates and densities that yield perfectly self-assembled single-domain magnetic crystals? What are the pathways that the liquid takes to crystallize? What role does the dipole strength play in the self-assembly and crystal structure? Can we seamlessly transit from a paramagnet to a single-domain ferromagnet and vice versa to create magnetic switches? The answers to such inquiries are relevant from the point of view of fundamental physics as well as technological applications.

We investigated some of these questions by cooling the SM fluid from the isotropic gas phase ($T_1>T_c$) to very low temperatures ($T_2<T_f$). The cooling rate is defined as $\Delta = (T_2-T_1)/N_T$, where $N_T$ is the number of MD steps in the quench. The rate is systematically decreased by increasing $N_T$. The main observations that emerge from this study are as follows: (i) For $\Delta < \Delta_c$, there is an onset of magnetic and spatial order at temperatures $T_m$ and $T_s$, respectively. Magnetic order sets first ($T_m>T_s$), characterized by the formation of long dipole-dipole chains. (ii) The crystalline order is established spontaneously and not by the traditional nucleation and growth mechanism. (iii) For the magnetic moment $\mu=1$, the colloidal self-assembly is a single domain face-centered crystal (FCC) with magnetization along the (111) direction. For the higher value of $\mu = 2.5$, the dipolar chains shrink due to strong attractive interactions. Consequently, the cubic colloidal crystal transforms into a body-centered orthorhombic (BCO) structure, and the magnetization is along the (001) direction. Self-assembled magnetic colloids can find extensive applications in various fields, including photonic crystals in display technologies and sensors \cite{Ge2008, Libaers2009}, magnetic switches \cite{Demirors2017}, energy storage systems \cite{Tian2022}, and data storage devices \cite{Bhushan2023}.

It is pertinent to discuss some earlier studies of equilibrated states that have addressed crystallization in magnetic fluids to highlight the significance of our observations. Experimental and theoretical studies of self-assembled monolayers have revealed ordered arrangements that are often polycrystalline or exhibit local defects in the form of vacancies and grain boundaries, compromising the quality of monolayers \cite{Swan2014, Kao2021, Tang2016, Zhang2020_}. To overcome these defect-related challenges and achieve nearly perfect crystalline configurations, various annealing protocols have been widely employed. These include thermal annealing, the application of external fields, and the incorporation of self-propelled particles \cite{Rajan2023, Kao2021, Van2016, Ramananarivo2019}. In $d=3$, self-assembly in magnetic colloids has been investigated using computer simulations. Groh and Dietrich, using density functional theory, obtained an FCC crystal as a stable structure \cite{Groh1996}. Klapp and Forstmann, on the other hand, minimized the free energy functional to find {\it body-centered tetragonal} (BCT) phase as the stable arrangement \cite{Klapp1997}. Monte Carlo (MC) simulations of Gao and Zeng for the highly dipolar fluid reveal BCO as a stable structure at low temperatures \cite{Gao2000}. Our investigations not only provide a systematic understanding of these diverse observations, but also provide protocols to obtain single-domain magnetic colloidal crystals for diverse applications.

Our paper is organized as follows. Sec.~\ref{sec2} provides the model and methodologies that are used to characterize magnetic and spatial order. The details of the simulations and numerical results are provided in Sec.~\ref{sec3}. Finally, in Sec.~\ref{sec4}, we end with a summary and discussion of our results.

\section{Model and Tools for Characterization}
\label{sec2}

\subsection{Stockmayer Model}

The SM model mimics magnetic fluids composed of spherical particles with magnetic moments embedded at their center. Consider $N$ identical particles with magnetic moment ${\vec{\mu}} = \mu\hat{\mu}$. The SM potential between particles $i$ and $j$ separated by a distance $\vec{r}_{ij} = r_{ij}\hat{r}_{ij}$ is given by \cite{Leeuwen1993, Stevens1995}: 
\begin{eqnarray}
\label{SM}
U(\vec{r_{ij}}, \hat{\mu_i}, \hat{\mu_j}) = 4\epsilon\sum_{i,j}\bigg[{\bigg( \frac{\sigma}{r_{ij}}\bigg)}^{12}-{\bigg( \frac{\sigma}{r_{ij}}\bigg)}^6\bigg]
+\frac{\mu_0 \mu^2}{4\pi}\sum_{i,j}\bigg[ \frac{\hat{\mu}_i.\hat{\mu}_j - 3(\hat{\mu}_i.\hat{r}_{ij})(\hat{\mu}_j.\hat{r}_{ij})}{r_{ij}^3} \bigg].
\end{eqnarray}
where $\hat{\mu}_i$ is the unit dipole vector of the $i^{th}$ particle. The first part corresponds to the LJ potential, with a short-ranged steric repulsion that prevents particle overlap and a weak van der Waals attraction that aids condensation. The parameters $\sigma$ and $\epsilon$ are the diameter of the particles and the depth of the attractive potential, respectively. They set the spatial and energy scales in the system. The second part represents the dipole-dipole interactions, which are anisotropic and significant up to large distances.

When cooled below the critical temperature ($T_c$), the SM fluid exhibits a phase separation from a gas phase to a gas-liquid co-existence phase. The phase diagram in the $\rho-T$ plane is conventionally obtained from the equation of state, but computer simulations provide an alternative route, especially when interactions are complex. This gas-liquid co-existence region has been extensively studied by many groups using MC and MD simulations and is known to occur for all values of dipole moments $\mu$ \cite{Van1993, Smit1989, Leeuwen1993, Bartke2007, Kalyuzhnyi2007, Dudowicz2004, Leeuwen1994, Hentschke2007}. The primary effect of $\mu$ is to increase the value of $T_c$, thus enlarging the coexistence region of the GL. Interestingly, studies reveal that the condensates assume density-dependent shapes with unusual arrangements of the magnetic moments \cite{Singh2023accelerated, Singh2023phase}.   

\subsection{Tools for Characterization}

The self-assembled structures of the SM gas exhibit both magnetic and spatial order. In the following, we discuss a variety of tools that are required to understand the consequences of these dual features.

\subsubsection{\textbf{Magnetization and Edwards-Anderson Order Parameter}}

At temperatures above the freezing temperature $T_f$, the dipolar particles form long chains in the coexistence region, leading to the formation of a large domain and substantial magnetic order in self-assembly. The resulting magnetization is the vector sum of all the moments:
\begin{equation}
\bf{M} = \frac{1}{N} \sum_{i=1}^{N} \hat{\mu}_i.
\end{equation}
The perfect ferromagnetic order corresponds to $M=1$, while the disordered or paramagnetic state corresponds to $M=0$. At lower temperatures $(T<T_f)$, the domains are smaller and randomly oriented due to freezing of moments. An appropriate order parameter for capturing the local arrangements of
dipole moments within the frozen morphologies is the Edwards-Anderson (EA)  order parameter defined as \cite{Edwards1975, Parisi1983}:
\begin{equation}
q_{EA}=\left[\langle \mu_i \rangle ^2\right]_{av},
\label{qEA}
\end{equation}
where $\langle...\rangle$ is the dynamical average that yields a non-zero value for frozen dipolar particles and $[...]_{av}$ is an ensemble average. In the paramagnetic phase, $q_{EA}=0$ and $M=0$. In the ferromagnetic phase, $q_{EA}\ne0$ and $M\ne0$. In the frozen (glassy) phase, on the other hand, $q_{EA}\ne0$ but $M\simeq0$.

\subsubsection{\textbf{Pair Correlation Function}}
\label{sec:PCF}

The standard probe to envisage the internal arrangements of particles within the condensed phase is the pair correlation function. This measures the probability of finding two molecules separated by distance $r$ relative to that of an ideal gas: $g(r) = \langle \overline{\rho(r)}\rangle/\rho_0$, where $\rho_0=N/V$ is the density of the ideal gas and $\overline{\rho(r)}$ is the average density of the system around $r$. The numerical evaluation is facilitated by the following formula \cite{Weis1993, Birdi2022}:
\begin{equation} \label{PCF}
g(r)=\frac{1}{N\rho_0}\bigg\langle\sum_{\stackrel{i,j=1}{i\neq j}}^N\frac{\delta(r-r_{ij})}{(4/3)\pi[(r+\Delta r)^3-r^3]}\bigg\rangle.
\end{equation}
The $\delta$- function is unity if $r_{ij}$ falls within the shell centered on $r$ and is zero otherwise. Dividing by $N$ ensures that $g(r)$ is normalized to a function per particle. By construction, $g(r)=1$ for an ideal gas, and any deviation implies correlations between the particles due to the inter-particle interactions. In the liquid phase, $g(r)$ exhibits a large peak at small $r$ signifying nearest-neighbor correlations followed by small oscillations that eventually approach 1 at large $r$. (The latter value represents the absence of correlation at large distance). The solid phase is characterized by several sharp peaks at values of $r$ that correspond to the lattice spacing of the crystal structures. 

The development of magnetization in the SM fluid introduces directional anisotropy along ${\bf M}$. An appropriate evaluation in this context is the directional pair correlation function \cite{Weis1993}:
\begin{align}
g_{\parallel}(r_{\parallel})=\frac{1}{N\rho_0}\bigg{\langle}\sum_{\stackrel{i,j=1}{i\neq j}}^N\frac{\delta(r_{\parallel} - r_{ij, \parallel})\theta(\sigma/2-r_{ij,  \bot})}{\pi (\sigma/2)^2 h}\bigg{\rangle}, 
\label{gpar}
\end{align}
where $ r_{ij, \parallel}= |\vec{r}_{ij}.\hat{M}| $ is the separation of particles along ${\bf M}$ and $ r_{ij,  \bot} = |\vec{r}_{ij}- (\vec{r}_{ij}.\hat{M})\hat{M}| $ is the separation in the perpendicular direction. The step function $\theta(x)$ ensures that the cylinder of radius $r=\sigma /2$ has a height $h$ that is used for the discretization of the simulation box. The pair correlation function in the perpendicular direction, $g_{\bot}(r_{\bot})$, can be evaluated analogously.

 \subsubsection{\textbf{Bond Order Parameter}}
\label{sec:BOP}

The local crystalline order in undercooled liquids and solids can be conveniently obtained using the local bond order parameters (BOP) $q_4$ and $q_6$ evaluated from \cite{Steinhardt1983, Errington2003, Shrivastav2021, Gasser2014}:
\begin{align}
\label{q1}
q_l(i)=\sqrt{\frac{4\pi}{2l+1}\sum_{m=-l}^{l}|\Bar{q}_{lm}(i)|^2},
\end{align}
with
\begin{align}
\label{q2}
\Bar{q}_{lm}(i)=\frac{1}{N_n(i)+1}\sum_{k=0}^{N_n(i)}q_{lm}(k),
\end{align}
and 
\begin{align}
\label{q3}
q_{lm}(k)=\frac{1}{N_b(k)}\sum_{j=1}^{N_b(k)}Y_{lm}(r_{kj}).
\end{align}
In Eq.~(\ref{q2}), $N_n(i)$ denotes all nearest neighbors of particle $i$ plus the particle $i$. In Eq.~(\ref{q3}), $N_b(i)$ denotes only the nearest neighbors of $i$. The nearest neighbors are identified as those that lie within a cut-off distance, which is determined by the first minimum of the pair correlation function. $Y_{lm}(r_{ij})$'s are the spherical harmonics, with $l$ as a free integer parameter and $m =-l,\cdot\cdot,l$. The BOPs or $ q_l(i)$'s have characteristic values for different structures and are indicated in Table~\ref{q}. 

\begin{table}[H]
\caption{Values of $q_4$ and $q_6$ for standard lattice structures.}
\begin{ruledtabular}
\begin{tabular}{ccc}
Structures & $q_4$ & $q_6$ \\
\hline
Simple cubic (SC) & 0.764 & 0.354 \\
Body-centered cubic (BCC) & 0.509 & 0.629 \\
Hexagonal close-packed (HCP) & 0.097 & 0.484 \\
Face centered cubic (FCC) & 0.191 & 0.575 \\
Body centered orthorhombic (BCO) & 0.200 & 0.566 
\end{tabular}
\end{ruledtabular}
\label{q}
\end{table}

\section{Simulation Details and Numerical Results}
\label{sec3}

MD simulations to study low-temperature morphologies of the SM fluid were implemented using LAMMPS software \cite{LAMMPS}. A system of $N=2400$ magnetic colloidal particles is taken within a cubic box of length $L=20$, which corresponds to a density $\rho=0.3$. We use the Ewald summation technique \cite{Frenkel2001} to account for (a) the long-range nature of the dipole-dipole interactions and (b) the tail of the LJ potential. The simulations were performed in the canonical ensemble (NVT) using periodic boundary conditions with a minimum image convention. The No\'se-Hoover thermostat (NHT) was used to correctly maintain the temperature and to preserve the relevant features of hydrodynamics for domain growth \cite{nose1984, hoover1985, hoover1999}.  For numerical integration, velocity-Verlet algorithms were used with simulation time step $\Delta t=0.002$ \cite{Velocity-verlet1983}. All calculations were performed in reduced LJ units by defining $ T^*=k_BT/\epsilon $, $ \rho^*=N\sigma^3/{V} $, $ \mu^*=\mu/\sqrt{\epsilon \sigma^3} $, $ \Delta t^*=\Delta t/\sqrt{m\sigma^3/\epsilon}$. (The star is dropped in subsequent discussions.) The data were averaged over 40 independent samples. 

\begin{figure}
\includegraphics[width=1\textwidth]{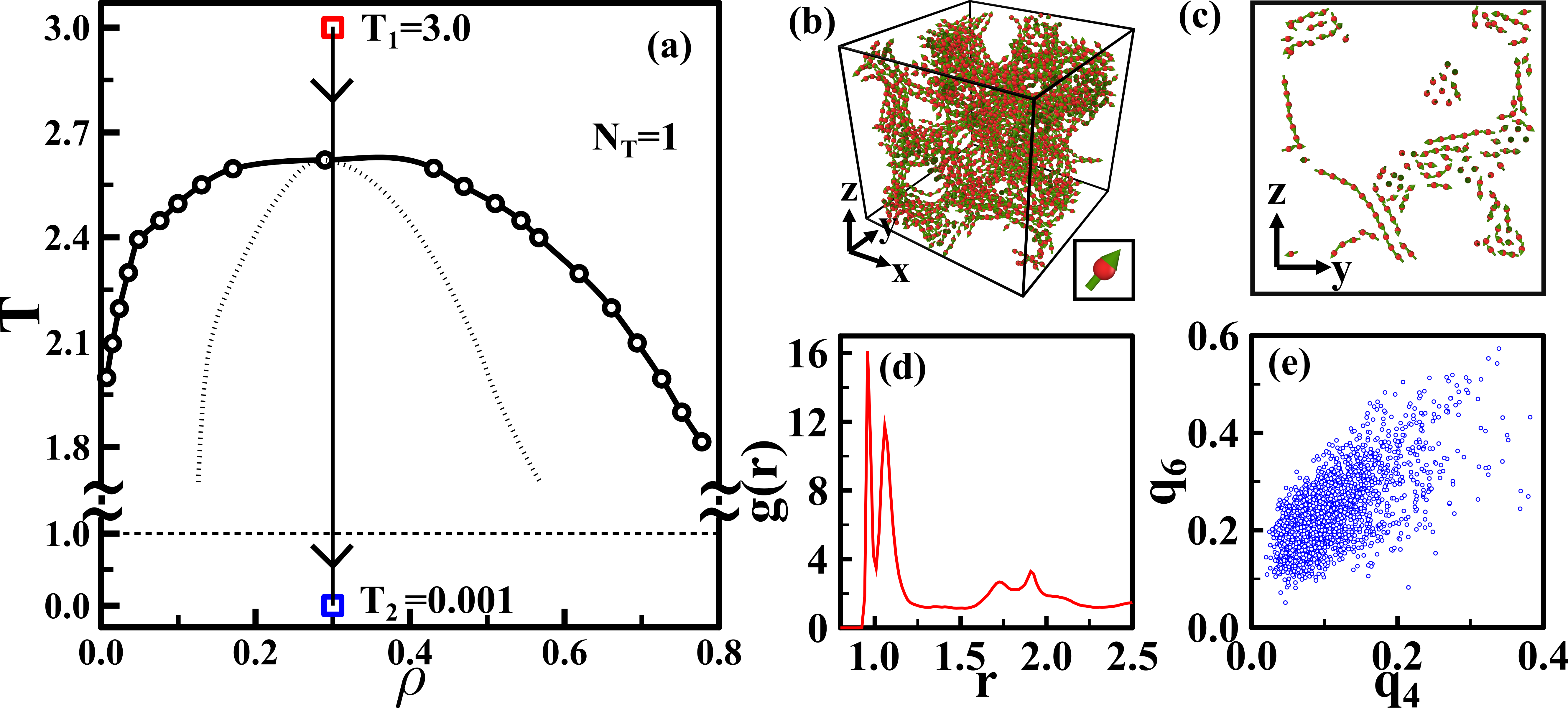}
\caption{(a) The GL coexistence phase diagram of the SM fluid with a dipole moment of $\mu=2.5$, read from Ref.~\cite{Stevens1995}. The solid line indicates the binodal region, while the dotted line separates the nucleation region from the spinodal region; see ref. \cite{Singh2023phase}. The dashed line separates the liquid phase from the solid phase. The arrow depicts the temperature quench from $T_1=3.0$ (red square) to $T_2 = 0.001$ (blue square). (b) Typical frozen morphology at $\rho=0.3$ after a quench $T_1\rightarrow T_2$ ($N_T=1$). A prototypical SM particle is also displayed for clarity. The morphology is characterized by an Edwards-Anderson order parameter $q_{EA}=0.97$ and a resulting magnetization of $M=0.001$. (c) The corresponding $yz$-slice from the center of the simulation box is also shown for visualization of local domains. (d) Pair correlation function $g(r)$ vs. $r$ of the frozen morphology in (b). (e) Scatter plot of the local bond order parameters $q_4$ and $q_6$.}
\label{phase}
\end{figure}

As a first exercise to understand the low temperature frozen morphology of the SM fluid with $\mu=2.5$, we start with a homogeneous SM gas of density $\rho=0.3$, equilibrated at high temperature $T_1=3$ (red square) above the GL coexistence curve indicated by the solid line in Fig.~\ref{phase}(a). The data \cite{Stevens1995} report a critical point at $\rho_c = 0.29(1)$, $T_c=2.63(1)$. The dotted curve separates the regions corresponding to phase separation via the mechanism of ``nucleation and growth'' and ``spinodal decomposition''. The system is cooled to $T_2$ in $N_T/(\Delta t) = 500~N_T$ MD steps, with a temperature reduction of $(T_1-T_2)/(500 N_T)$ after each MD step. First, consider an instantaneous quench ($N_T=1$) to $T_2=0.001$ in the spinodal region as shown in the GL phase diagram in Fig.~\ref{phase}(a). The resulting morphology (after equilibrating for $10^6$ MD time steps) is shown in sub-figure (b). For clarity, the corresponding $yz$-slice is also shown in sub-figure (c). There are small domains emerging from the co-alignment of short chains, and these are randomly oriented. An evaluation of magnetization and the Edwards-Anderson order parameter yields $M \simeq 0$ and $q_{EA} \simeq 1$, indicating the formation of an amorphous glassy state that lacks positional order. This is corroborated by the evaluation of $g(r)$ vs. $r$ in sub-figure (d), which does not show nearest-neighbor peaks corresponding to any known crystal structure. The evaluation of the local bond order parameter (for each particle) is shown in sub-figure (e). The spread in the values of $q_4$ and $q_6$ clearly indicates a distinct neighborhood for each particle, indicating non-crystallinity.

\begin{figure}
\centering
\includegraphics[width=0.8\linewidth]{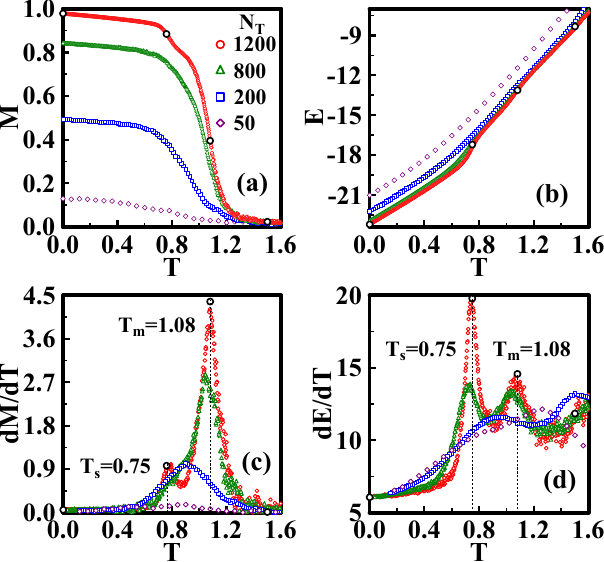}
\caption{(a) Development of magnetization $M$ as the SM fluid cools from the homogeneous state ($T_1=3.0$) to the solid state ($T_2=0.001$) at a density of $\rho=0.3$ for different time steps $N_T$=1200, 800, 200, and 50. Corresponding plots of (b) total energy $E$ vs. $T$, (c) $dM/dT$, and (d) $dE/dT$. The points of inflection occur at $T_m=1.08$ and $T_s=0.75$. The open circles in the sub-figures indicate data points that will be explored later.}
\label{mag_T}
\end{figure}

Next, we study the role of $N_T$. An increase in $N_T$ corresponds to a slower quench. Figs.~\ref{mag_T}(a) and (b) show the magnetization $M$ and the energy $E$ as a function of the temperature $T$ for four different values of $N_T =$ 50, 200, 800, 1200. As seen in Fig.~\ref{mag_T}(a), the magnetization builds up from $M=0$ (paramagnetic state) to non-zero values as the cooling rate becomes slower. In particular, $M\rightarrow1$ for $N_T = 1200$, indicating a co-alignment of dipoles and the formation of a single domain. As seen in Fig.~\ref{mag_T}(b), this also corresponds to the state of lowest energy. To capture the magnetic transition, we also show the variation of (c) $dM/dT$ vs. $T$ and (d) $dE/dT$ vs. $T$. For lower values of $N_T=50$, 200, 800, there is a point of inflection at $T_m(N_T)$. For $N_T=1200$, the slow cooling rate captures two points of inflection at $T_m=1.08$ and $T_s=0.75$. The pronounced peak at $T_m \simeq 1.08$ is indicative of a para $\rightarrow$ ferro transition, while the peak at $T_s \simeq 0.75$ signals the liquid $\rightarrow$ solid transition. We have checked that $T_m$ and $T_s$ vary $\lesssim 10^{-5}$ for $N_T=1600$. The corresponding evaluations of $dE/dT$ vs. $T$ in (d) also corroborate the above observations. 

\begin{figure}
\centering
\includegraphics[width=0.8\linewidth]{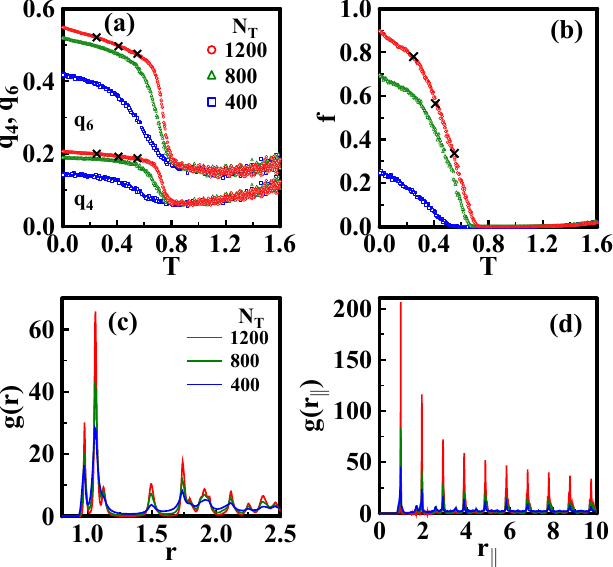}
\caption{(a) The development of local bond order parameters $q_4$ and $q_6$ as a function of dynamic temperature $T$ for $N_T=$1200, 800, and 400. (b) The corresponding fraction of particles with $q_6 \geq 0.5$ indicates crystalline order. Corresponding plots of (c) pair correlation function $g(r)$ vs. $r$ and (d)  directional pair correlation function $g(r_\parallel)$ vs. $r_\parallel$ along the magnetization $\bf {M}$. The crosses in (a) and (b) mark morphologies that will be explored later.}
\label{crys_T}
\end{figure}

In Fig.~\ref{crys_T}, we perform a set of evaluations to check the spatial organization of the dipoles in the domains. The sub-figure (a) shows $q_4$, $q_6$ vs. $T$ for $N_T =$ 400, 800 and 1200. The lowest temperature values of $q_4$ and $q_6$ are highly dependent on $N_T$. For $N_T=1200$, these values are $q_4 \simeq 0.19$ and $q_6\simeq0.56$, suggesting FCC or BCO structure; see Table \ref{q}.  The sub-figure (b) shows the fraction of particles $f$ vs. $T$ that exhibit crystalline order. Clearly, the fraction increases with slow cooling. We find that while most initial states for $N_T=1200$ exhibit crystalline order at quench temperature $T_2=0.001$, approximately 10$\%$ of the initial states are metastable with a few frozen sites. We also evaluate in sub-figure (c), $g(r)$ vs. $r$ for the morphologies obtained at $T_2$ after each of the cooling protocols. The peaks indicate the presence of periodicity, and the peak positions indicate the distances between neighbors. The lattice structure and the corresponding unit cell can be obtained from the separation between the consecutive peaks. We will address these evaluations shortly.  In addition, $g_{\parallel}(r_{\parallel})$ vs. $r_{\parallel}$ along ${\bf M}$ is also shown in sub-figure (d). There are sharp equidistant peaks that suggest the formation of layered dipole chains in this direction.

\begin{figure}
\centering
\includegraphics[width=1\linewidth]{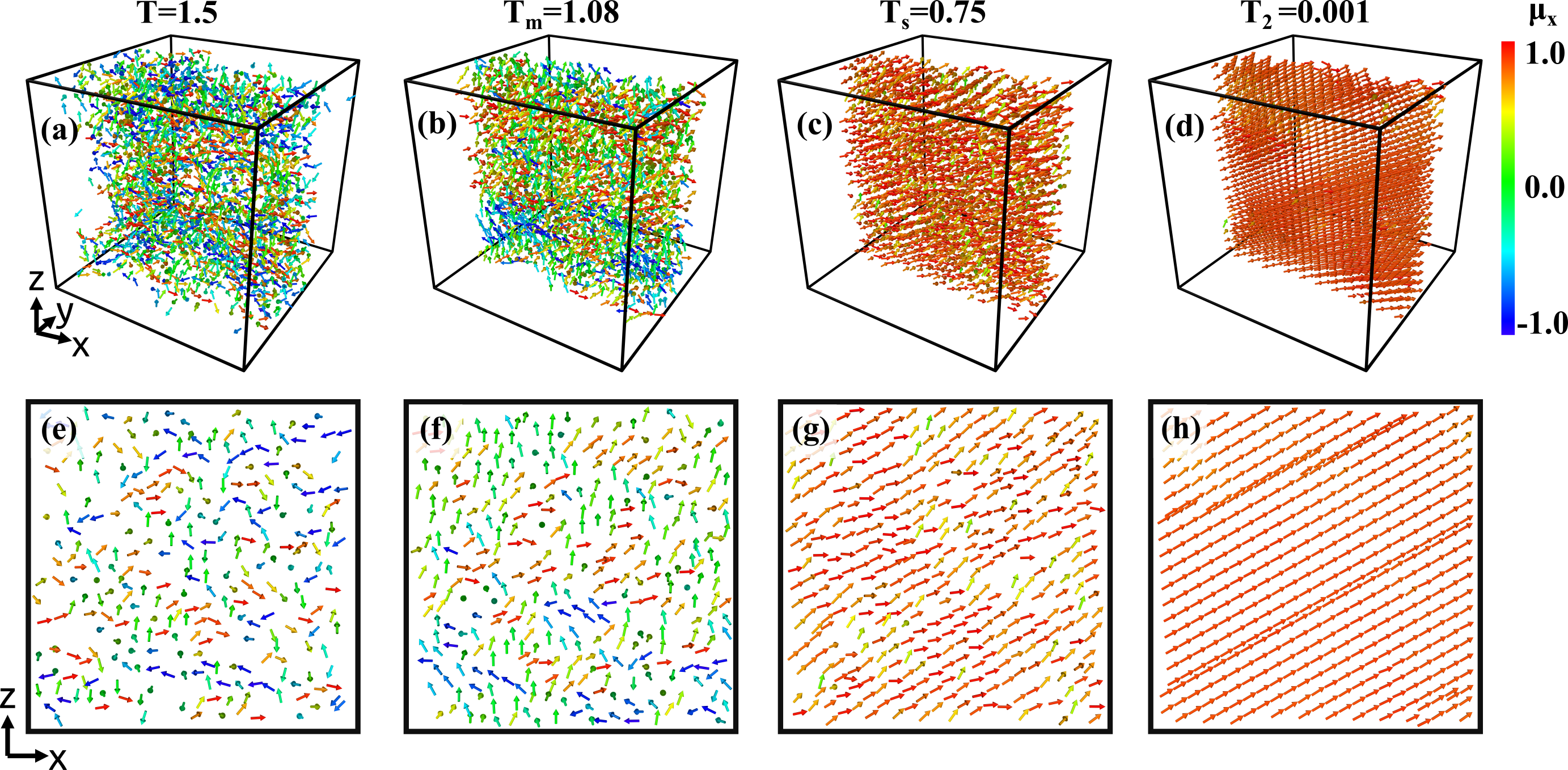}
\caption{Top row: Evolution morphologies for $N_T=1200$ at various temperatures: (a) $T=1.5$, (b) $T_m=1.08$, (c) $T_s=0.75$, and (d) $T_2=0.001$ that are indicated by open circles in Fig.~2. The color code indicates the orientation of the dipole moment along ${\bf M}$. Bottom row: Corresponding $xz$-slices at the center of the morphologies are displayed in (e)–(h) to clearly depict the alignment of dipolar vectors.}
\label{mag_mor}
\end{figure}

\begin{figure}
\centering
\includegraphics[width=0.8\linewidth]{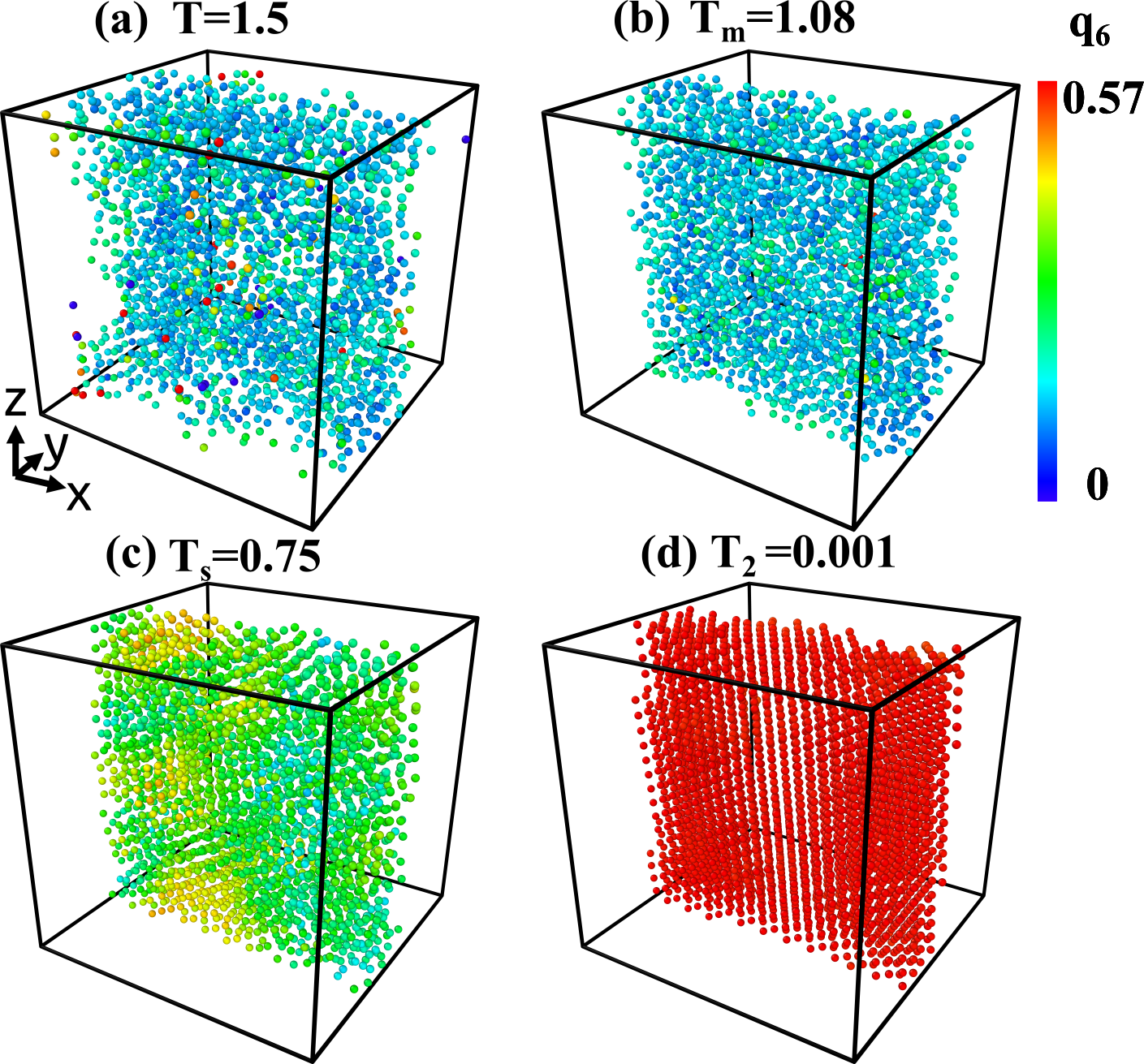}
\caption{Same morphologies as in Fig.~\ref{mag_mor}. Here, the color code indicates the value of the local bond order parameter $q_6$.}
\label{q6_mor}
\end{figure}

Clearer physical insights can be obtained by studying the morphologies in Fig.~\ref{mag_mor} at four characteristic temperatures indicated by the open circles in Fig.~\ref{mag_T}: (a) $T=1.5$, (b) $T_m=1.08$, (c) $T_s=0.75$ and (d) $T_2=0.001$. For clarity, we also show the corresponding vertical slices in sub-figures (e)-(h). The color coding indicates the orientation of the dipole moment along the average direction of magnetization ${\bf M}$. In sub-figure (a), the system is still a paramagnet with a small fraction of the condensed phase. The magnetic order starts to build up in sub-figure (b), corresponding to the first point of inflection at $T_m=1.08$ due to dipole-dipole interactions. As the temperature is reduced to $T_s$, the morphology in (c) reveals the formation of short dipole chains and the emergence of the preferred direction $\hat{M}$.  At $T_2=0.001$, as seen in snapshots (d) and (h), the chains span the system. The morphology eventually evolves into a single domain (as $T\rightarrow T_2$), comprising long chains of dipoles that are coaligned. Thus, the point of inflection at $T_m$ signals the onset of ferromagnetic order from a disordered paramagnetic state in the SM model. There is also an increase in the value of the bond order parameter $q_6$ when the SM gas cools slowly. This can be observed in Figs.~\ref{q6_mor}(a)-(d) for the corresponding temperatures, the color code now representing the value of $q_6$. Magnetic order triggers the development of crystalline order.

\begin{figure}
\begin{center}
\includegraphics[width=1\linewidth]{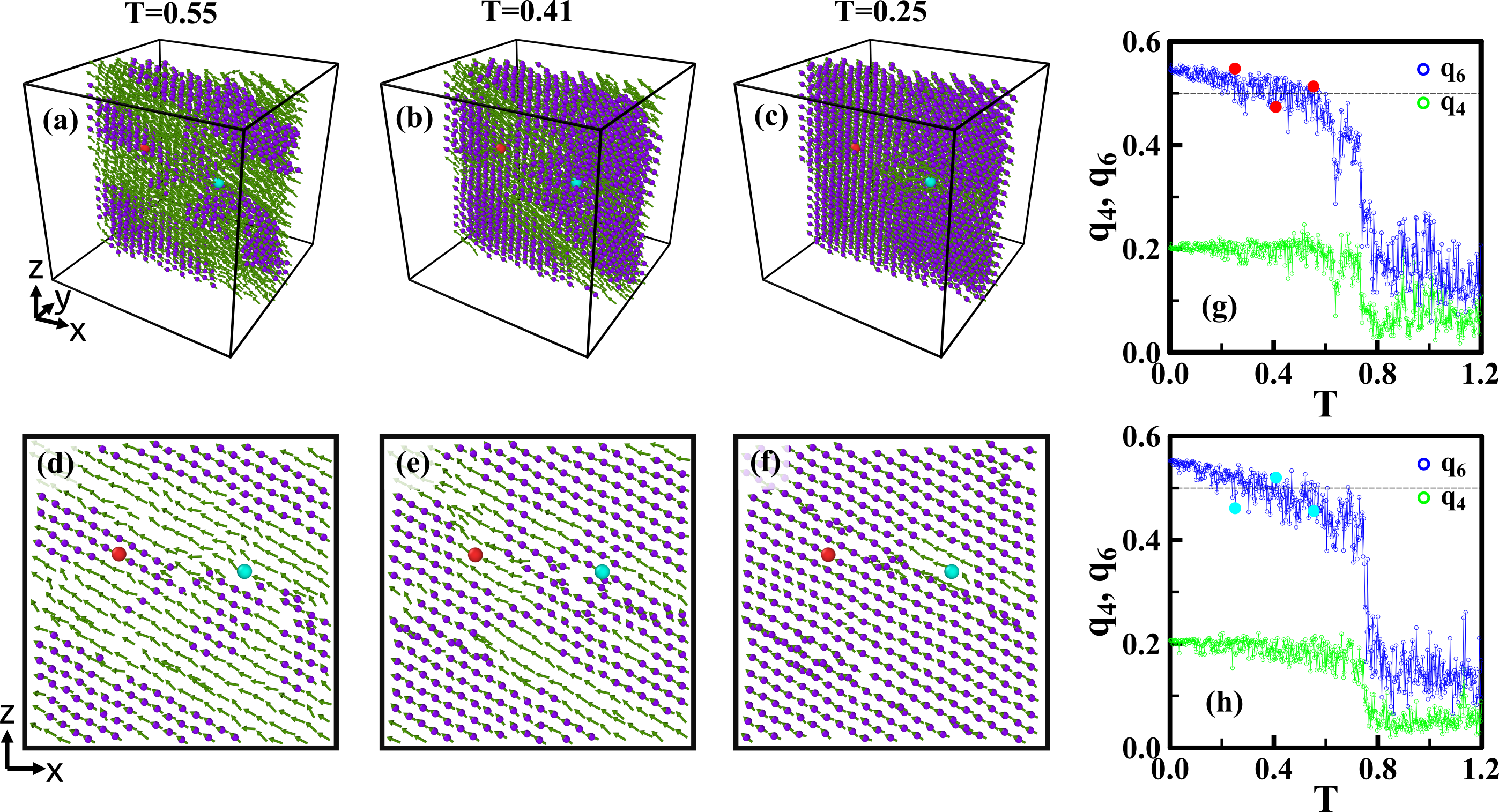}
\end{center}
\caption{Typical evolution of morphologies and tracking of two prototypical particles indicated by red and cyan colors at $N_T=1200$ for temperatures (a) $T=0.55$, (b) $T=0.41$, and (c) $T=0.25$. (These values of temperature  are indicated by crosses in Figs.~\ref{crys_T}(a) and (b).) Particles with $q_6 \geq 0.5$ are marked by purple dots, while the green arrows indicate the direction of their dipole moment. The corresponding xz-slices are provided in (d)-(f) for clearer visualization. The evolution of $q_4$ and $q_6$ with $T$ is shown in (g) for the red particle and in (h) for the cyan particle.  The dotted horizontal line at $q_6=0.5$ indicates the reference value for crystalline order.} 
\label{crys_mor}
\end{figure}

We further explore the development of the crystalline order. In Fig.~\ref{crys_mor}, we track the local crystalline environment of two archetypal particles, colored red and cyan, as the temperature is lowered from $T_1\rightarrow T_2$. They are indicated in the morphologies for three representative values of temperature: (a) $T = 0.55$, (b) $T=0.41$ and (c) $T=0.25$. (These values of $T$ are indicated by crosses in Fig.~\ref{crys_T}.) The corresponding $xz$-slices are also shown in sub-figures (d) - (f). The evolution of $q_4$ and $q_6$ with $T$ is shown in (g) for the red particle and (h) for the cyan particle. Clearly, the spatial environment around the particles is dynamic, implying that crystal growth does not occur through nucleation. Instead, the transition from a non-crystalline state to a crystalline state emerges simultaneously at several locations uniformly throughout the volume, indicating that there are no thermodynamic barriers to be overcome. We infer that this phase change occurs spontaneously, as in {\it spinodal decomposition}.

\begin{figure}
\begin{center}
\includegraphics[width=0.9\linewidth]{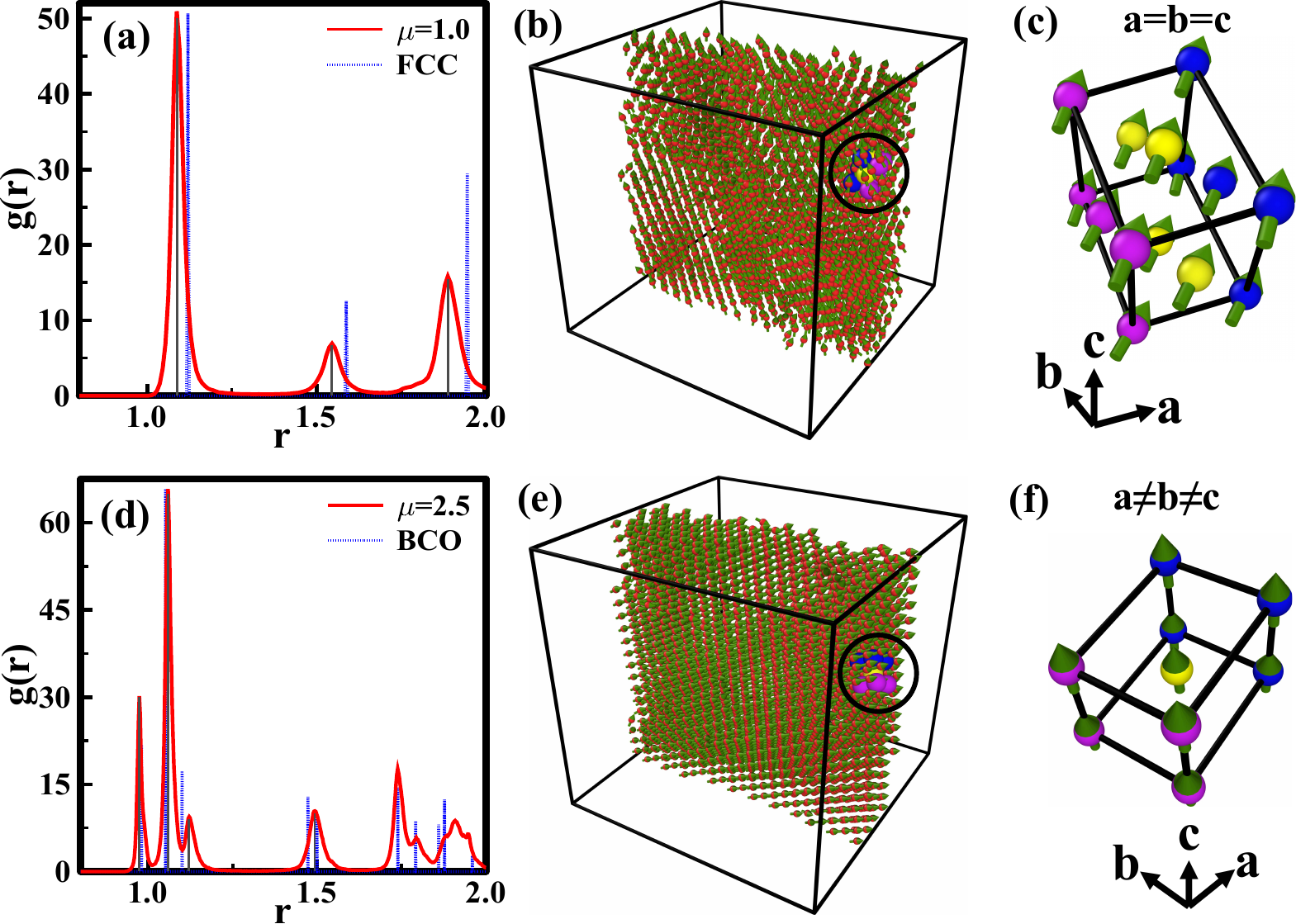}
\end{center}
\caption{(a) Pair correlation function $g(r)$ vs. $r$ at $N_T=6000$, $T=0.001$, and $\mu=1.0$ is shown by the solid (red) line. The dotted (blue) curve is the corresponding plot for an ideal FCC crystal structure. The thin vertical lines indicate the $r$ values that have been used for the evaluation of the lattice parameters. (b) A prototypical morphology showing a unit cell constructed using nearest neighbor values read from the pair correlation function (thin vertical lines in (a)). (c) Enlarged view reveals an FCC structure with $a=1.54(1)$. Magenta particles represent the layer at $x=0$, yellow particles represent the layer at $x=a/2$, and blue particles represent the layer at $x=a$. The orientation of dipole moments represented by green arrows is along (111). (d) $g(r)$ vs. $r$ at $N_T=1200$ for $\mu=2.5$ (solid red line) and the ideal BCO crystal structure (dotted blue line). (e) A prototypical morphology with the unit cell constructed using nearest neighbor $r$ values (indicated by thin vertical lines in (d)). (f) Enlarged BCO unit cell having $a=1.51(1)$, $b=1.12(1)$, $c=0.97(1)$, where magenta particles represent the layer at $x=0$, yellow particles the layer at $x=a/2$, and blue particles the layer at $x=a$. The magnetization is along (001).} 
\label{mu}
\end{figure}

We study the effect of the dipole moment on the crystal structure in Fig.~\ref{mu}. For precision, we used morphologies that are obtained after slow cooling with $N_T=6000$ to avoid freezing even a small number of dipoles. The red curve in sub-figure (a) shows the sample-averaged $g(r)$ vs. $r$ for $\mu=1$. The peak values at $r=$ 1.09, 1.54 and 1.89 correspond to the first, second and third nearest neighbors. The blue curve is the pair correlation function for a perfect face-centered cubic (FCC) lattice ($a=b=c$) with neighbors at $a/\sqrt{2}$, $a$, $\sqrt{3/2}a$ and is provided for reference. A prototypical morphology obtained from our simulations at the end of the slow quench is shown in sub-figure (b). We extract the unit cell of this ordered structure using the nearest-neighbor distances read from the thin lines in $g(r)$ vs. $r$. This reconstruction, shown in sub-figure (c), is undoubtedly an FCC unit cell with the lattice spacing evaluated as $a=1.54(1)$. The corresponding density is $\rho_{\rm FCC} = 4/a^3 = 1.095$. The sites corresponding to the three layers are colored magenta, yellow, and blue for clarity. The orientation of dipole moments is along the (111) direction. A similar exercise was conducted for the $\mu=0$ LJ fluid. We also observed that the resulting self-assembly also exhibits an FCC structure. The identification of FCC structures in weakly dipolar fluids is in agreement with the findings of Wang et al. \cite{Wang2013}, who determined through MD simulations that FCC structures are observed for reduced dipole moments up to $\sqrt{3}$. This observation is further corroborated by the work of Groh and Dietrich \cite{Groh1996}, who employed density functional theory in their analysis.

The self-assembly for $\mu=2.5$ is shown in the second row of Fig.~\ref{mu}. The blue curve in the sub-figure (d) corresponds to an ideal BCO structure $(a\ne b \ne c)$ with the nearest neighbors at $c$, $\sqrt{a^2+b^2+c^2}/2$, $b$, $\sqrt{b^2+c^2}$. The peak values 0.97, 1.06, 1.12, and 1.49 of $g(r)$ are used to extract the corresponding unit cell. The latter, shown in sub-figure (f), is BCO with $a=1.51(1)$, $b=1.12(1)$, and $c=0.97(1)$. The corresponding density is $\rho_{\rm BCO} = 1.219$. Note that the dipole-dipole interactions favor head-to-tail alignment, which results in the formation of long chains of magnetic moments. For $\mu=2.5$, there is further compression of the chains due to stronger dipole-dipole interactions. Consequently, there is a transformation from the cubic FCC ($a=b=c$) to the non-cubic BCO $(a\ne b \ne c$) structure. Further, the orientation of dipole moments is now along the (001) direction. The observed BCO crystal structure for strong dipolar fluids is consistent with the ground-state energy ($T=0$) calculations of Groh and Dietrich \cite{Groh2001}. (We have also investigated quenches in the nucleation region where the aggregates are neither single-domain nor crystalline.) 

\begin{figure}
\centering
\includegraphics[width=0.8\linewidth]{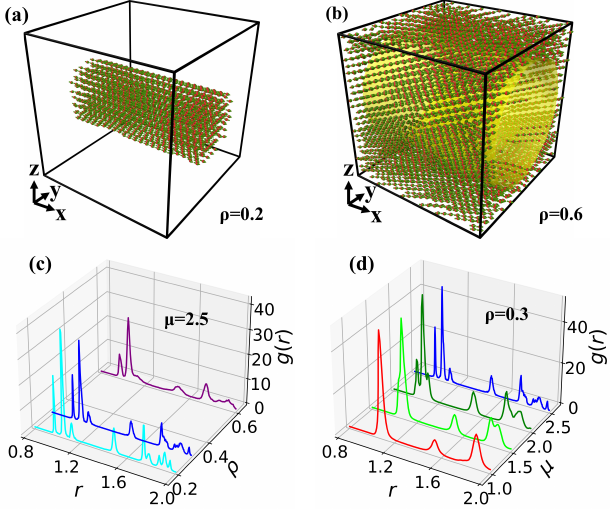}
\caption{Aggregates formed by particles with $\mu=2.5$ particles after cooling to $T=0.001$ at densities (a) $\rho=0.2$ and (b) $\rho=0.6$. The yellow color in (b) indicates a hollow (cylindrical) region. (c) Corresponding pair correlation function $g(r)$ vs. $r$.  is illustrated against particle separations for densities $\rho=0.2$, 0.3, and 0.6, evaluated at dipole strength $\mu=2.5$. (d) Pair correlation functions evaluated from aggregates formed at $T=0.001$ and $\rho=0.3$ for a range of dipole moments $\mu=1.0$, 1.5, 2.0, and 2.5.}
\label{rho_mu}
\end{figure}

Finally, in Fig.~\ref{rho_mu}, we explore how changes in $\rho$ and $\mu$ affect the formation of single-domain magnetic crystals. The aggregates obtained when a gas with $\mu=2.5$ is cooled to $T=0.001$ are shown in Figs.~\ref{rho_mu}(a) ($\rho = 0.2$) and \ref{rho_mu}(b) ($\rho = 0.6$). They are single domain with density-dependent shapes, as observed in our earlier work \cite{Singh2023phase}. Both morphologies exhibit magnetic order along the long axis. The internal arrangements of the dipolar particles are investigated by evaluating $g(r)$ vs. $r$, shown in Fig.~\ref{rho_mu}(c) with the $\rho=0.3$ data for reference. The nearest-neighbor peaks of $g(r)$ are density independent, and the condensates corresponding to all three densities have a BCO structure. Next, we investigate how changes in $\mu$ affect the internal structure. Fig.~\ref{rho_mu}(d) shows $g(r)$ vs. $r$ for aggregates formed by cooling the gas with $\rho = 0.3$ to $T=0.001$ with $\mu=1.0,1.5,2.0,2.5$. The particle separation undergoes a continuous change from FCC ($\mu=1.0$) to BCO ($\mu=2.5$) as the dipolar strength increases.

\section{Summary and Discussion}
\label{sec4}

We have addressed the formation of colloidal magnetic crystals, a topic of great interest in recent years. Our starting point is a gas of dipoles at high temperatures $(T_1>T_c)$ that is cooled to very low temperatures ($T_2<T_f$) in the solid state. The quench from $T_1\rightarrow T_2$ is interrupted several times. We find that for a critical cooling rate, the system of paramagnetic dipoles with $\mu = 1$ forms a FCC structure. The magnetic moments align to form chains along the $(111)$ direction, thus producing a single-domain magnetic colloidal crystal. For large $N_T$ (slow quenches), the onset of magnetic order ($T\simeq 1$) precedes the onset of crystalline order ($T\simeq 0.8$) highlighting that magnetic order drives the crystallization transition. The strength of $\mu$ also plays a role in the crystalline order. We observe a structural transition from FCC$\rightarrow$BCO as the dipolar strength increases. Further, single-domain colloidal crystals (FCCs or BCOs) are seen only for near-critical quenches, indicating that they are important for complete spatial ordering. 

The phase space of the model parameters is large for the SM model. Our investigations provide only a flavor of the promising phenomena in this magnetic fluid. However, the methodologies and frameworks that we have developed set the stage for many stimulating investigations. For example, understanding the two phase transitions that signal the onset of magnetic and spatial order is of fundamental interest. The possibility of identifying newer lattice structures by a systematic variation of the magnetic moment or manipulating the attractive and repulsive terms in the SM model can also be exciting. Tailored multi-domain aggregates in the nucleation region, the incorporation of magnetic fields, and the possibility of thermal and magnetic hysteresis may be valuable for applications such as magnetic switches. 

Colloidal self-assembly is also emerging as a promising and practical approach to the fabrication of photonic structures such as crystals, glasses, and opals \cite{Kim2011}. Their diffraction properties can be easily manipulated by tuning parameters such as density, particle size, attractive and repulsive interactions, etc. Magnetically responsive colloidal photonic crystals greatly expand the scope of technological applications due to the possibility of using magnetic fields in addition to incident radiation \cite{Ge2008}. Our work demonstrates the versatility of aggregated colloidal structures that emerge from the interplay of cooling protocols, magnetic moments, and particle density. We hope that this will be a step towards establishing synergy between theoretical modeling, experimental investigations, and technological applications in this contemporary field.

\bibliographystyle{apsrev4-1}
\bibliography{biblio.bib}

\end{document}